\documentclass[reprint,amsmath,aps,amssymb,prl,showpacs,superscriptaddress]{revtex4-1}
\usepackage{bmpsize}
\usepackage{graphicx}	
\usepackage[dvipsnames]{xcolor}
\usepackage{amssymb}
\usepackage{natbib}
\usepackage{array}

\newcommand{\gillespeaks}[1]{{\color{black} #1}}
\usepackage[normalem]{ulem}
\newcommand{\DEf}{\Delta E_{\rm eff}}

\begin{document}

\title{\textcolor{black}{From glass formation to icosahedral ordering by curving three-dimensional space}}
\author{Francesco Turci}
\email[Corresponding author: ]{f.turci@bristol.ac.uk}
\affiliation{H.H. Wills Physics Laboratory, Tyndall Avenue, Bristol, BS8 1TL, UK}
\author{Gilles Tarjus}
\affiliation{LPTMC, CNRS-UMR 7600, Universit\'e Pierre et Marie Curie, bo\^ite 121, 4 Pl. Jussieu, 75252 Paris cedex 05, France}
\author{C. Patrick Royall}
\affiliation{H.H. Wills Physics Laboratory, Tyndall Avenue, Bristol, BS8 1TL, UK}
\affiliation{School of Chemistry, University of Bristol, Cantock's Close, Bristol, BS8 1TS, UK}
\affiliation{Centre for Nanoscience and Quantum Information, Tyndall Avenue, Bristol, BS8 1FD, UK}
\affiliation{Department of Chemical Engineering, Kyoto University, Kyoto 615-8510, Japan}

\begin{abstract}
Geometric frustration describes the inability of a local molecular arrangement, such as icosahedra found in metallic glasses and in model atomic glass-formers, to tile space. Local icosahedral order however is strongly frustrated in Euclidean space, which obscures any causal relationship with the observed dynamical slowdown. Here we relieve frustration in a model glass-forming liquid by curving 3-dimensional space onto the surface of a 4-dimensional hypersphere. For sufficient curvature, frustration vanishes and the liquid ``freezes'' in a fully icosahedral structure via a sharp ``transition''. Frustration increases upon reducing the curvature, and the transition to the icosahedral state smoothens while glassy dynamics emerge. %s. 
Decreasing the curvature leads to decoupling between dynamical and structural length scales and the decrease of kinetic fragility. This 
%{connects} 
sheds light on the observed glass-forming behavior in %the 
Euclidean space.
%{frustrated} Euclidean space {to the transition in curved space.}
 \end{abstract}

\pacs{61.20.-p; 64.70.Q-; 61.20.Ja}

\maketitle

%\textit{Introduction --- } 
The very large increase in viscosity found in glass-forming liquids upon cooling or compression without significant change in structure remains a major outstanding challenge in condensed-matter physics. 
In particular, one seeks to clarify whether vitrification is linked to an underlying thermodynamic phase transition or whether the process is predominantly dynamical \cite{berthier2011}.

Among the most enduring pictures of dynamic arrest is that liquids form geometric motifs upon supercooling \cite{frank1952}. It is now possible to identify such motifs {, {\it e.g.}, icosahedra and {other} %related 
\emph{locally preferred structures} (LPS),} using computer simulation \cite{%steinhardt1983,jonsson1988,dzugutov2002,ganesh2006,
coslovich2007,charbonneau2012,hocky2014,royall2015physrep} and particle-resolved studies in colloid and granular experiments \cite{royall2008,%mazoyer2011,
leocmach2012}. Further evidence of local icosahedral order is %upon cooling is also 
found in metallic glass-formers \cite{sheng2006,shen2009,cheng2011,liu2013,hirata2013}. {While the idea of icosahedra (to focus on this specific LPS) as being the cause of dynamical slowdown in many materials has proven to be remarkably durable, it has, equally remarkably, seldom been seriously tested. This is the main motivation of the present work.}

\gillespeaks{A strong piece of evidence for a structural or thermodynamic mechanism would be the identification of static length scales that grow significantly when approaching the glass transition. (It is indeed possible to demonstrate that the divergence of the relaxation time at a finite temperature implies a divergent static correlation length \cite{montanari2006}, but this relies on a bound that may not necessarily put stringent constraints in the dynamically accessible regime.)  Any successful theory also  needs to account for the well-established phenomenon of dynamical heterogeneities, in the form of ``liquid-like'' fast-moving and ``solid-like'' slow-moving regions whose lifetime and size increase upon supercooling \cite{ediger2000,berthier2011dynamical}, a phenomenon that has been characterized by ``dynamical'' length scales. Different types of static lengths have been considered in previous studies \cite{biroli2008,berthier2012,hocky2012,charbonneau2012,karmakar2014}, and here we focus on the case of static lengths related to icosahedral order and their coupling/decoupling to dynamical length scales.}

{Five-fold symmetric motifs such as icosahedra 
%\cite{frank1952,dzugutov2002,coslovich2007,leocmach2012,charbonneau2012,malins2013jcp}. 
do not tile 3d Euclidean space periodically \cite{frank1952}. %, which may then suppress crystallization This property has been further developed in the framework of geometric frustration.} 
For single-component systems of spheres, it has been theoretically shown \cite{coxeter,sadoc2006} and observed in simulations \cite{straley1984} that $120$ particles on the surface of a 4d hypersphere, the ``3-sphere'' $S^3$, of a specific curvature %(in units of the particle radius) 
can realize a perfect tiling of space with every particle at the center of an icosahedron: the so-called $\{3,3,5\}$ polytope. Flattening space then induces frustration \cite{sadoc2006,nelson2002}. However, at the end of the flattening process, in Euclidean space, frustration is strong and the growth of icosahedral order is strongly suppressed \cite{charbonneau2012,royall2015physrep}. In particular, {for multi-component mixtures of spheres} at the degree of supercooling accessible to computer simulations (and colloid experiments), \textit{i.e.}, the first 4-5 decades of increase of the structural relaxation time $\tau_\alpha$ relative to the normal liquid, rather limited domain sizes of icosahedral regions are found and the associated length scales remain small \cite{coslovich2007,malins2013jcp,royall2015physrep}. Furthermore, these structural lengths are significantly smaller than {some }%the 
dynamical lengths associated with the growingly heterogeneous character of the dynamics \cite{charbonneau2012,charbonneau2013pre,charbonneau2013jcp,malins2013jcp}.

Despite {the} many claims {and} %or 
suggestions, this calls into question whether such structures can be the main cause of dynamic arrest. At the very least, it is fair to state that the description of the mechanism by which frustrated icosahedral order influences slow dynamics remains an unresolved problem. To make progress on this issue, we curve 3d space to relieve frustration and we use curvature as an additional control parameter to investigate equilibrium glass-forming liquids. 
{While curved 3d space can only be realized by computer simulations, it is a unique means to probe the causal link between local icosahedral order and dynamics and to test the premise upon which geometric frustration is predicated --- that of an underlying phase transition avoided due to frustration \cite{tarjus2005}.}

We consider a model glass-forming liquid, the Wahnstr\"om model, which is a Lennard-Jones binary mixture with size ratio $\sigma_A/\sigma_B=6/5$. {We choose this model because it} is known to display a significant correlation between slow dynamics and the formation of local icosahedral motifs in Euclidean space \cite{pedersen2010,malins2013jcp,hocky2014}. 

We perform Monte Carlo (MC) simulations of $N\in[120,720]$ particles on {the 3-sphere} $S^3$, with a modified Marsaglia method \cite{marsaglia1972,kratky1982} %in order 
to isotropically sample the surface of the 4-d hypersphere. The results in curved space are complemented with molecular dynamics (MD) results in Euclidean space \cite{MCvsMD}. We fix the reduced density
\begin{equation}
\tilde{\rho}=\dfrac{N}{V(R)}\dfrac{V_{\rm cap}(R, \sigma_A)+V_{\rm cap}(R, \sigma_B)}{E(\sigma_A)+E(\sigma_B)}=1.296\sigma_A^{-3},
\label{eq:density}
\end{equation}
where $N$ is the total number of particles, $V(R)$ the (hyper) area of the $3$-sphere, $V_{\rm cap}(R, \sigma)$ is the (hyper)area of a spherical cap of height $h=R(1-\cos{\sigma/2R})$ and $E(\sigma)$ the Euclidean volume of a particle of diameter $\sigma$. At fixed density the number of particles $N$ and the radius of curvature $R$ are therefore coupled: the range $N=120-720$ corresponds to $R\approx 1.666-3.037\sigma_A$. In the limit $R\to\infty$, one recovers the usual expression $\tilde{\rho}\to N/V$ (see the Supplementary Material (SM) for more details \cite{SM}).

\begin{figure}[t]
\centering
\includegraphics[scale=1]{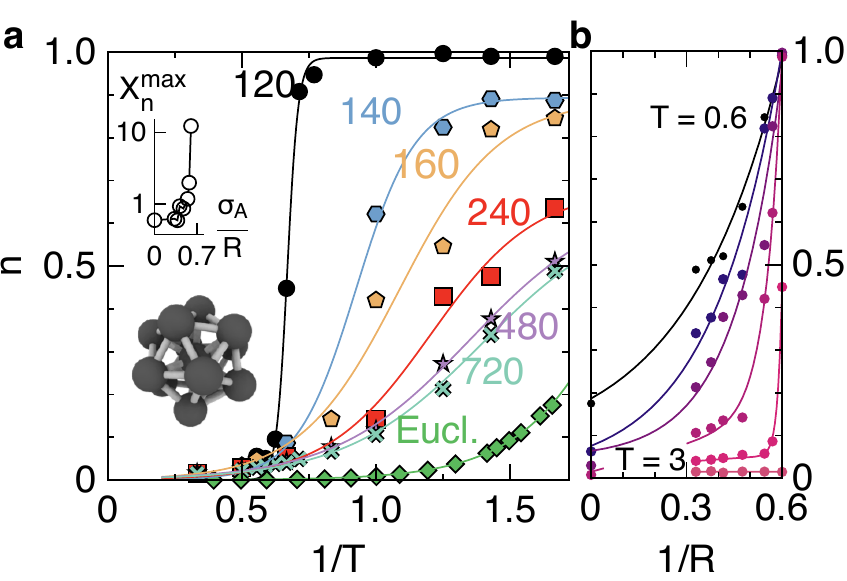}
\caption{ (a) Concentration $n$ of particles {found in domains of icosahedra (see ball-and-stick model)} as a function of the inverse of the reduced temperature $T$ for several curvatures characterized by the system size $N$ (see Eq.\ref{eq:density}). The lines are hyperbolic-tangent fits, from which the maximum of the derivative $\chi_{n}=dn/dT$ can be estimated: see the case $N=120$ in the inset. (b) Same data, as a function of the curvature $1/R$ at $T=0.6,0.7,0.8,2,3$. Lines are guides to the eye. }
\label{figNicos}
\end{figure}

We first investigate the effect of the curvature on the structure of the system. {We take the first minimum of the pair correlation function as the bond length which,} together with the Delaunay triangulation obtained from the convex hull of the particle coordinates, {provides } the network of nearest neighbors (see SM for more details). This then allows for the detection of icosahedral order throughout the system via a modified topological cluster classification \cite{malins2013tcc}.

We find that for $N=120$ the bidisperse Wahnstr\"om model abruptly freezes to an ordered icosahedral structure, the $\{3,3,5\}$ polytope \cite{coxeter},  as the temperature $T$ is lowered, just like a monodisperse system of spherical particles \cite{sadoc2006,straley1984}. This is illustrated in Fig.~\ref{figNicos} where we plot the concentration $n$ of particles detected in icosahedral domains as a function of $1/T$ for various curvatures characterized by the total number of particles $N$. For $N=120$ a sharp crossover, which is the finite-size version of a first-order transition, from an icosahedra-poor liquid to an icosahedral structure is {found}. Frustration is thus relieved by curvature and the concentration fluctuations due to the bidispersity have no significant influence at this curvature. {(See SM for a more detailed analysis of the low-$T$ structure% is given in the SM
.)}

As curvature decreases (and $N$ and $R$ increase), the crossover smoothens: the growth of icosahedral order becomes more gradual while the maximum concentration of icosahedra saturates at lower values, which is a sign of increasing frustration. The temperature range over which the change takes place broadens and shifts to lower temperatures. The Euclidean case is the end point of this continuous variation with curvature {(see also the SM)}.

To describe the slowing down of the dynamics while avoiding the complexity brought by curvature and the parallel transport along geodesics we consider a simple time-dependent correlation function {based on the number of neighbors which are lost with time}:
%derived from the definition of the neighborhood:
\begin{equation}
C(t)=\left\langle \dfrac{1}{N}\sum_{i=1}^N\dfrac{ \vec{v}_i(t_0+t)\cdot \vec{v}_i(t_0) }{\vec{v}_i^2(t_0)}\right\rangle_{t_0},
\end{equation}
where $\vec{v}_i(t)$ is the indicator vector of length $N$ identifying the nearest neighbors of particle $i$ at time $t$. The function $C(t)$ corresponds to the average fraction of neighbors that has not changed between time $t_0$ and time $t$. While being independent of the local curvature of the space, it provides a measure of the %slow ($\alpha$) 
relaxation. Through a stretched-exponential fit to $C(t)-C(\infty)$ (see SM) we obtain an estimate of the structural relaxation time $\tau$. In the case of the two larger curvatures, $N=120$ and $140$, the crossover is so sharp that the relaxation time jumps from a finite value to an exceedingly large one in the icosahedral state, which then behaves as a solid {for our purposes}. This is much like the dynamical behavior at a first-order transition, albeit here in a finite-size system: the relaxation time %does not 
{may not }truly diverge but is too large to be accessible in a computer simulation. In contrast for $N=160$, the crossover is smooth enough that we can access the relaxation time even when the growth of icosahedral order has saturated and we then see no sign of divergence.

The %results for the 
relaxation time {is} {shown} in Fig.~ \ref{figTaus} (a). For $T\gtrsim 2$ curvature has little %or no 
influence on the relaxation (see also the SM). But this is no longer true at lower temperature. While the two largest curvatures {exhibit} an abrupt freezing to a solid icosahedral phase, the transition appears to be \textit{avoided} for weaker curvatures and a continuous increase of the relaxation time is found over the accessible range, as in the Euclidean space.

In order to assess the change with curvature of the kinetic fragility, \textit{i.e.}, the degree of super-Arrhenius temperature dependence of the relaxation time, we consider the effective activation energy $\DEf=T \log(\tau/\tau_{\infty})$ where $\tau_{\infty}$ is the relaxation time at high $T$ %,: it is {shown} in 
[Fig.~\ref{figTaus} (b)]. The two curvatures where freezing takes place behave very differently from the others. For $N\geq 160$ to the Euclidean limit, $\DEf$ {is found to}
increase continuously with increasing $1/T$, which is the signature of a super-Arrhenius, fragile, behavior. The differences between the curvatures are not dramatic but there is a clear trend towards a monotonic decrease of fragility as curvature decreases. Since the high-$T$ behavior is independent of curvature, this can be seen unambiguously and without data-fitting by comparing the effective activation energies (or the relaxation times) at low $T$ (see Fig.~\ref{figTaus}): The kinetic fragility decreases as the curvature decreases (and at the same time frustration increases, consistent with previous work \cite{kawasaki2007, sausset2008}).

\begin{figure}[tp]
\centering
\includegraphics[scale=1]{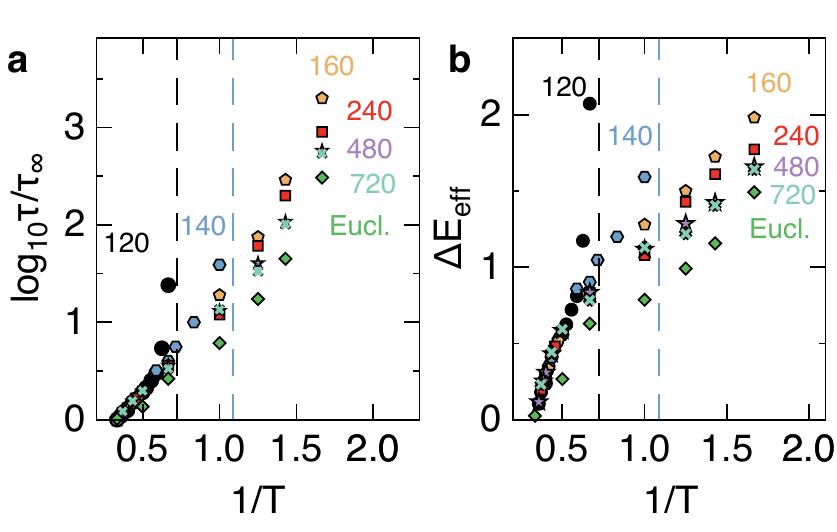}
\caption{(a) Logarithm of the relaxation time (a) and Effective activation energy $\DEf=T \log \tau/\tau_{\infty}$ (b) versus $1/T$ for several curvatures. The vertical dashed lines approximately indicate the temperatures at which the $N=120$ and $N=140$ systems freeze into a solid icosahedral phase.}
\label{figTaus}
\end{figure}

%\textit{Correlations and length scales ---} 
As mentioned above, the emergence of slow dynamics in glass-forming systems is often attributed to the growth of spatial correlations in the dynamics and the statics \cite{karmakar2014}. The former manifest themselves as dynamical heterogeneities \cite{berthier2011dynamical}; the latter are found through {investigations of point-to-set correlations  \cite{%bouchaud2004,
montanari2006,%cavagna2007,
biroli2008,berthier2012,hocky2012,charbonneau2012} }or through some 
characterization of the growth of the local 
%{amorphous} 
order \cite{royall2015physrep,karmakar2014,cubuk2015}. As also already emphasised, for most glass-formers studied by %computer 
simulations, including the Wahnstr\"om mixture, one finds a rapid increase for the dynamical lengths but a modest increase of the static lengths \cite{charbonneau2013pre,malins2013jcp%,dunleavy2015
}. One is %of course 
limited by the dynamic range accessible to computer simulations, so that {it is hard to attain} the deeply supercooled regime near the glass transition. %In addition, 
Thus, it is hard to clearly identify on the origin of the observed decoupling. We cannot improve the accessible range but we can add a new control parameter, the curvature.

In order to explore dynamic correlations, we focus on low-mobility (slow) particles, {following} \cite{Flenner2011}. To do so, we define a neighbor-dependent mobility and use a thresholded persistence function of the indicator neighbor vectors $v_i$ in order to identify the slow particles. The number of slow particles is then defined as 
\begin{equation}
	N_{\rm slow}(t)= \left\langle\sum_{i=1}^N\Theta[\vec{v}_i(t_0+t)\cdot \vec{v}_i(t_0)-\tilde{N}]\right\rangle_{t_0}
\end{equation}
where $\Theta(x)$ is the Heaviside function and $\tilde{N}$ the minimum number of neighbors of a particle that must not change for this particle to be taken as slow: we chose $\tilde{N}=8$ but we checked that the results are not very sensitive to the choice of this particular threshold ($5\leq \tilde{N} \leq 10$). We can then study the average of the number of slow particles during time $t$ and the fluctuations, characterized by the susceptibility $\chi(t)=(1/N)(\langle N_{\rm slow}^2(t)\rangle-\langle N_{\rm slow}(t)\rangle^2)$.

\begin{figure}[t ]
\begin{center}
\includegraphics[]{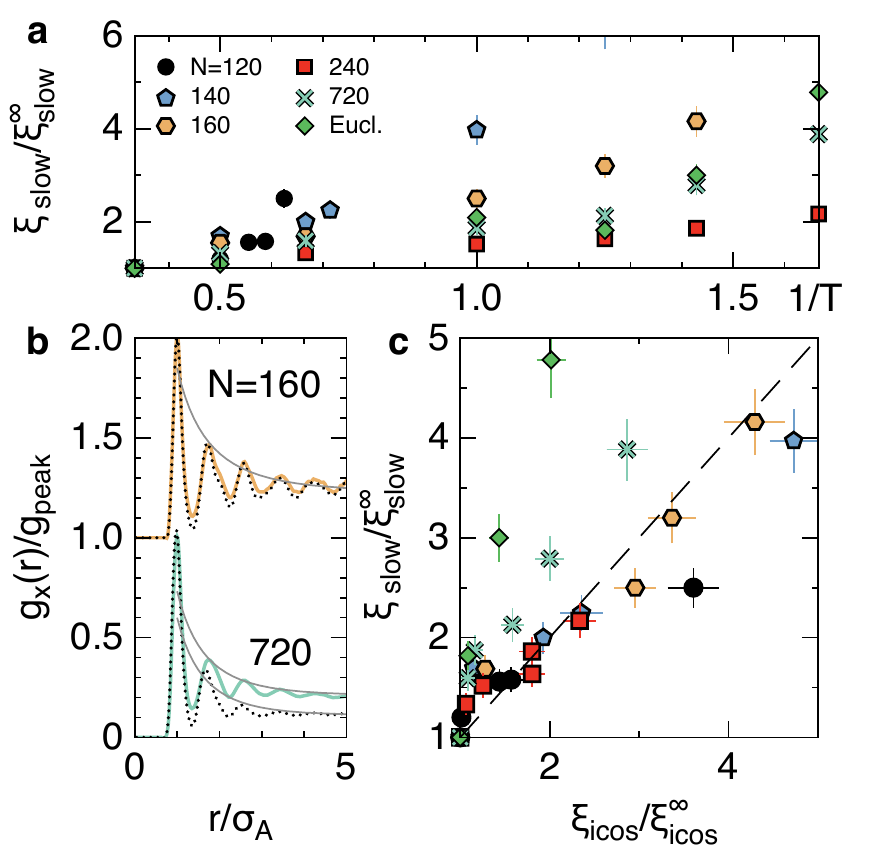}
\caption{(a) {Rescaled dynamic length versus $1/T$ for different curvatures.} 
%Temperature dependence of the maximum $\chi^{\rm max}$ of the dynamical susceptibility $\chi(t)$. In the inset: $\chi(t)$ for $N=240$ at $T=1.5,0.8,0.7$. 
(b) Radial distribution functions $g_{\rm slow}(r;\tau)$ (continuous) and $g_{\rm icos}(r)$ (dotted) for $N=160,720$ and $T=0.8$ \gillespeaks{scaled by the first peak height and shifted for clarity. Fits to extract lengthscales (see SM) in grey}. (c) Rescaled dynamic length versus the rescaled structural length for different curvatures, down to the Euclidean limit. }
\label{figLengths}
\end{center}
\end{figure}

To extract the dynamic length, we work in real space \cite{smallsystems}: we compute the radial distribution function restricted to the particles that are slow at $t=\tau$, $g_{\rm slow}(r;\tau)$. From it we estimate a typical correlation length {$\xi_\mathrm{slow}$} via an exponential fit, $g_{\rm slow}(r;\tau)\sim \exp[-r/\xi_{\rm slow}(\tau)]/r+c$, where $c$ is a long-range normalization constant {depending on} the finite-size limitations of our systems: see Fig.~\ref{figLengths} (b) and the SM. The resulting length, after a rescaling by its high-temperature value, is shown in Fig.~\ref{figLengths}(a) for several curvatures. It grows as $T$ decreases, which indicates increasing spatial correlations in the dynamics and bigger dynamical heterogeneities. {The rate of change with $T$ appears nonmonotonic with curvature, first decreasing with $N$ down to $N=240$ and then increasing up to the Euclidean limit. 
%(with again a markedly different behavior for $N=120,140$). 
(We find a less marked but} qualitatively similar behavior for the peak value $\chi^{\rm max}$ of the dynamic susceptibility $\chi(t)$, which occurs for $t\approx\tau$ as generically found in glass-formers and can loosely be taken as a relative measure of the number of dynamically correlated particles{%\cite{berthier2005}}
: {see %the 
SM.)}

To obtain a structural length scale, we use a similar approach to that for the dynamic %one, 
{length, } except that we consider only particles in icosahedra: we compute the corresponding restricted radial distribution function $g_{\rm icos}( r)$ and extract $\xi_\mathrm{icos}$ through an exponential fit [see Fig.~\ref{figLengths}(b) and the SM]. 
%The same comment as before is in order for $N=120,140$ and one finds 
We find a {steady} reduction of both the extent and the {rate} (with decreasing temperature) of the growth of icosahedral correlations as curvature decreases and frustration increases, 
in line with the results shown in Fig.~\ref{figNicos}(a).

The dynamic and structural lengths are compared in Fig.~\ref{figLengths}(c), once rescaled to their high-temperature value. One observes a clear trend with increasing curvature ({\it i.e.}, decreasing $N$): while a significant decoupling is found in %the 
Euclidean space, this decoupling decreases and appears to vanish for $N=240$ and less. When the icosahedral order becomes less frustrated, dynamical and structural lengths {increase} %go 
hand in hand as the relaxation slows down. The growth of the local order then seems to fully determine the properties of the dynamics. On the other hand, as frustration increases, this one-to-one correspondence {becomes} blurred and other mechanisms, possibly related to the mean-field description of glass-forming liquids \cite{kirkpatrick1989,charbonneau2016}, must be considered in addition. Note that we do not expect the decoupling to be {merely an} effect of the finite size of the curved systems. It has indeed been shown (in Euclidean space) that in the %\paddyspeaks{dynamical} 
range accessible to %dynamical 
simulations the dynamics of 3d glass-formers is not very sensitive to size effects {\cite{karmakar2009,
berthier2012finite}, %quite 
contrary to 2d systems \cite{flenner2015,vivek2016long}.

%\textit{Conclusion ---} 
To summarize: We have studied the structure and the dynamics of a supercooled liquid in curved 3d space, using curvature as a way to tune the degree of frustration of the local order. {Through this {\it additional control parameter} one can assess the {\it causal} relationship between local order and glass formation.}
%shed light on the otherwise limited information one has on the role of local order in glass formation. 
%, more specifically on the role of icosahedral order in 3d systems such as the Wahnstr\"om model.
Evidence for some correlation between {relaxation slowdown 
%the slowing down of relaxation and the 
and growth of icosahedral order has been reported  in the present Wahnstr\"om model} {\cite{hocky2014,malins2013jcp,coslovich2007}}, but it is hard to get an in-depth picture considering the limited range accessible to simulations and the strong frustration. Starting from the Euclidean limit and curving space, we find 
%a continuous evolution with 
an increase of the extent and of the influence of the local icosahedral order on %the physics of 
the liquid under cooling, 
%including an increase of the kinetic fragility, 
until one encounters a low enough frustration that allows freezing {into} an ordered icosahedral structure and thereby prevents glass formation.

Interestingly, the increase of frustration with decreasing curvature is accompanied by the decoupling of the temperature evolution of the dynamical and structural lengths. This suggests that while the collective behavior of the system is controlled by the growth of the icosahedral order and the proximity to an underlying (avoided) ordering transition for sufficiently weak frustration, the slowing down is no longer uniquely dominated by the local order when frustration increases: the observed decoupling appears as a signal that other mechanisms come into play. The behavior found in the Euclidean space is the end point of this process with only remnants of the role of icosahedral ordering.

{Based on Fig.~\ref{figNicos},  we speculate that at deep supercooling {in Euclidean space}, beyond the regime accessible here, the population of icosahedra would increase and ultimately plateau. 
% as we have found for the smaller $N=120-160$ systems. 
This behavior may be accessible to advanced experimental techniques such as nanobeam electron diffraction \cite{hirata2013,liu2013}. 
%It is further possible that 
Such a saturation 
%of structural order 
might also herald a fragile-to-strong crossover in metallic glass-formers known for their icosahedral order \cite{zhou2015}.}

Finally, we contrast the situations in {$d=3$ and $d=2$.} The change of behavior with curvature observed in the present study is profoundly different from that found in 2d systems 
%of spherical particles, 
where $6$-fold local bond-orientational order is prevalent. In the latter case, the ordering transition in the absence of frustration (which 
%in 2d 
means in the Euclidean plane) is continuous or weakly first-order \cite{nelson1979dislocation,bernard2011}. In 3d, the transition appears strongly first-order: it is accordingly characterized not by the continuous divergence of the relaxation time or the correlation length but by (rounded) jumps from finite to  {exceedingly large} values in these quantities. 
%, jumps that are moreover rounded by the intrinsic finite size of the systems
On the other hand, by curving 2d space \cite{sausset2008,sausset2010,vest2014dynamics} one then encounters an \textit{avoided continuous transition} near which the correlation length can be very large. Here instead, by flattening 3d space, we see the effect of an \textit{avoided first-order transition}, with a broadened crossover and limited correlation lengths. The collective {static} behavior generated by the proximity of an avoided 
%ordering 
transition is more prominent in 2d than in 3d. This may explain why the decoupling between dynamical and static lengths appears to be absent in many 2d liquids \cite{kawasaki2007,tanaka2010,sausset2010} and why finite-size effects are more dramatic in 2d than in 3d glass-formers \cite{flenner2015}. 
%On top of this, the influence of growing structural correlations on the relaxation slowdown may also be different: in 2d, the defects in the local $6$-fold order are point-like while in 3d icosahedral order one expects defect lines, whose dynamics may then be strongly constrained \cite{nelson}. This deserves further investigation. 

\textit{Acknowledgements---} We are grateful to S. Taylor for stimulating conversations about {algorithms in curved space}, to J. Hicks for preliminary simulations and to P. Charboneau for helpful suggestions. CPR acknowledges the Royal Society for funding and FT and CPR acknowledge the European Research Council (ERC consolidator grant NANOPRS, project number 617266). {CPR acknowledges the University of Kyoto SPIRITS fund.}
 This work was carried out using the computational facilities of the Advanced Computing Research Centre, University of Bristol.

\bibliography{mercifulReleaseLessCitations}

\end{document}